# Fabrication and characterization of pseudo-spin-MOSFET*

Y. Shuto[1,5], R. Nakane[2,5], H. Sukegawa[3,5], S. Yamamoto[4,5], M. Tanaka[2,5], K. Inomata[3,5], and S. Sugahara[1,5]

[1]*Imaging Science and Engineering Laboratory, Tokyo Institute of Technology*
[2]*Department of Electrical Engineering and Information Systems, The University of Tokyo*
[3]*Magnetic Materials Center, National Institute for Materials Science*
[4]*Department of Information Processing, Tokyo Institute of Technology.*
[5]*CREST, Japan Science and Technology Agency*

Recently spin-transistors receive considerable attention as a highly-functional building block of future integrated circuits. In order to realize spin-transistors, it is essential that technology of efficient spin injection/detection for semiconductor channel is established. However, this is not so easy challenge owing to ferromagnet/semiconductor-interface-related several problems. In this paper, we demonstrate pseudo-spin-MOSFET (PS-MOSFET) architecture that is a new circuit approach using an ordinary MOSFET and magnetic tunnel junction (MTJ) to reproduce the functions of spin transistors [1].

Figure 1 shows the circuit configuration of PS-MOSFET. A MTJ connected to the source of a MOSFET feeds back its voltage drop to the gate, and the degree of negative feedback depends on the resistance states of the MTJ. Therefore, the actual input bias $V_{GS0}$ and also substrate (body-source) bias $V_{BS0}$ can be varied by the magnetization configurations of the MTJ even under a constant gate bias ($V_G$) condition. Therefore, the PS-MOSFET can possess high and low current drivabilities that are controlled by the magnetization configurations of the MTJ, as shown in Fig 2(a). In addition, magnetic-field-free current-induced magnetization switching (CIMS) for the MTJ can be established by increasing $V_G$ (Fig. 2(b)). Thus, the PS-MOSFET can reproduce the spin-transistor behavior and would be the most promising spin-transistor based on present MRAM technology. Figure 2(c) shows output characteristics in the case that a MTJ is connected to the drain of an ordinary MOSFET. The differences in the drain currents between parallel and antiparallel magnetization configurations are tiny for this circuit configuration owing to the lack of the feedback effect.

In order to demonstrate spin-transistor action of PS-MOSFET, a prototype PS-MOSFET was fabricated using a MTJ with a full-Heusler alloy ($Co_2FeAl$; CFA) electrode and MgO tunnel barrier, as shown in Fig. 3. In our study, the bottom-gate MOSFET structure with a SOI substrate was employed for simplicity, in which the buried oxide (BOX) and Si substrate were used as a gate dielectrics and gate electrode, respectively. The fabrication procedures are as follows: A p-type 100-nm-thick (001) SOI wafer was used as a substrate. The SOI thickness was reduced to 20 nm by thermal oxidation of the SOI layer. The channel and source/drain regions were defined by the $SiO_2$ layer patterned as a hard mask, and then n-type impurity (P) was thermally doped into the source/drain region. The physical channel length and width of the fabricated PS-MOSFET were 10 μm and 110 μm, respectively.

A spin-valve-type MTJ was fabricated on the thermally-grown atomically-flat $SiO_2$ layer adjacent to the source region. Firstly, a Ru(7nm) / IrMn(12nm) / CoFe(3nm) / MgO(1.5nm) /CFA(20nm) multilayer was deposited by RF sputtering at room temperature on the $SiO_2$ layer using a 10-nm-thick MgO buffer layer. Subsequently, post-annealing treatment was performed at 300°C for quality improvement of ordering structure in the CFA film. After this treatment, the CFA film exhibited highly (001)-oriented and *B*2-ordered structure. During the post-annealing, a magnetic field was applied to the sample for sufficient exchange biasing. Then, the multilayer film was formed into a rectangular shape of 20 × 50 μm$^2$.

Figure 4 shows the resistance value of the fabricated MTJ in the antiparallel and parallel magnetization configurations as a function of bias voltage. The MTJ exhibited clear exchange-biased TMR characteristics and a relatively high tunneling magnetoresistance (TMR) ratio of 62 % at room temperature, as shown in Fig. 5. Figure 6 shows the output characteristics of the fabricated PS-MOSFET. Non-saturation behavior in the output characteristics can be attributed to source junction leakage and gate leakage currents that were caused by unoptimized ion milling process for the MTJ. However, the field-effect transistor behavior was clearly observed. The drain current ($I_D^P$) in parallel magnetization was higher than that ($I_D^{AP}$) in antiparallel magnetization, indicating that the PS-MOSFET can operate as a spin-transistor. Figure 7 shows the drain current as a function of magnetic field. The drain current well reflects the resistance change of the MTJ. A magnetocurrent ratio $\gamma_{MC}$ ($=(I_D^P-I_D^{AP})/I_D^{AP}$) of 15.2 % was achieved at drain bias $V_D$ of 0.5V and gate bias $V_G$ of 5V. Figure 8 shows $\gamma_{MC}$ as a function of $V_D$. $\gamma_{MC}$ increased with decreasing $V_D$ and increased with increasing $V_G$.

Nonvolatile logic devices such as nonvolatile flip-flop (NV-FF) and nonvolatile SRAM (NV-SRAM) are one of the most important applications for PS-MOSFETs. NV-SRAM can be simply configured by connecting two PS-MOSFETs to the storage nodes of a standard SRAM cell, as shown in Fig. 9. NV-FF can be also configured in the same manner. These nonvolatile devices cell can be shut down without losing its logic information [1]. PS-MOSFETs would be a key device for power-gating architecture.

**References**: [1] Y. Shuto, S. Yamamoto, and S. Sugahara, J. Appl. Phys., **105**, 07C933 (2009).
*Presented at Intl. Conf. on Silicon Nano Devices in 2030, Tokyo, October 13-14, 2009, pp. 148-149.

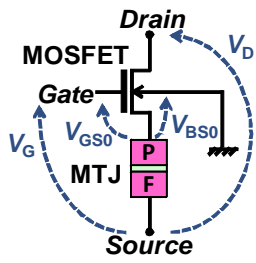
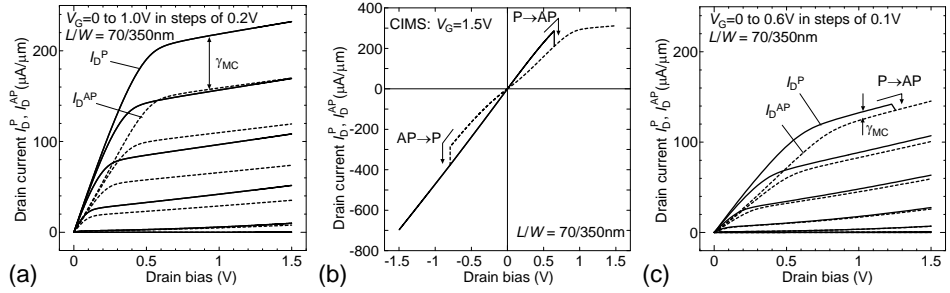

Fig. 1: Circuit configuration of proposed PS-MOSFET.

Fig. 2: (a) Simulated output characteristics of a PS-MOSFET with MTJ parameters of $R_P$=5k$\Omega$, TMR=100%, $V_C$=0.5V, and $V_{half}$=0.5V (whose notations are described in Ref. [1]). (b) CIMS behavior of the PS-MOSFET. (c) Output characteristics in the case that MTJ is connected to the drain of an ordinary MOSFET.

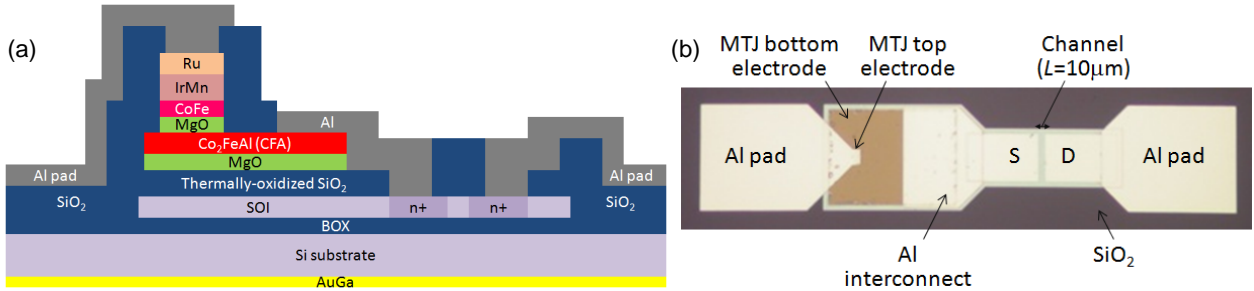

Fig. 3: (a) Schematic side view of a fabricated PS-MOSFET. (b) Photograph of the fabricated PS-MOSFET.

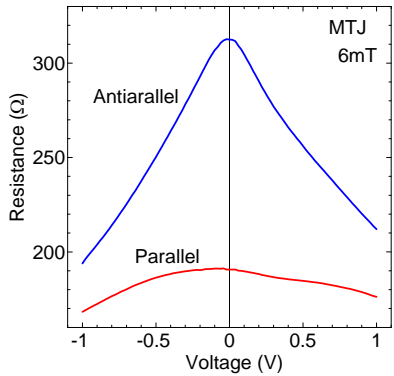
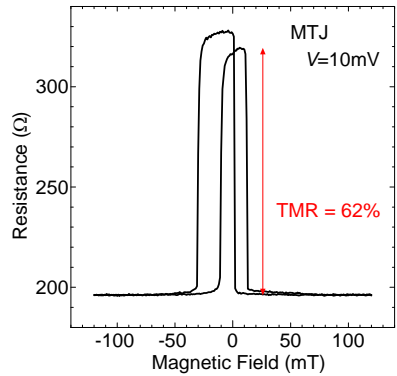
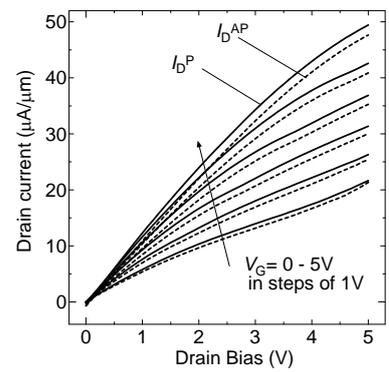

Fig. 4: Junction resistance as a function of applied voltage for the fabricated MTJ in the parallel and antiparallel magnetization configurations.

Fig. 5: Junction resistance as a function of magnetic field with applied bias voltage of 10mV.

Fig. 6: Output characteristics of the fabricated PS-MOSFET. The drain currents are plotted as a function of drain bias $V_D$, where gate bias $V_G$ varies from 0 to 5 V in steps of 1V. Solid and broken curves show the drain currents in the parallel and antiparallel magnetization configurations, respectively.

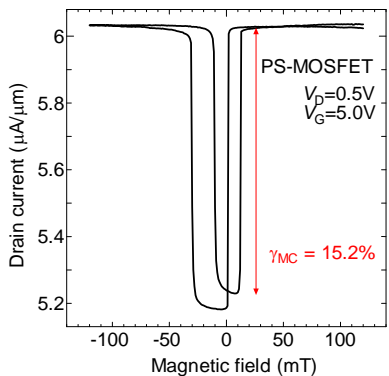
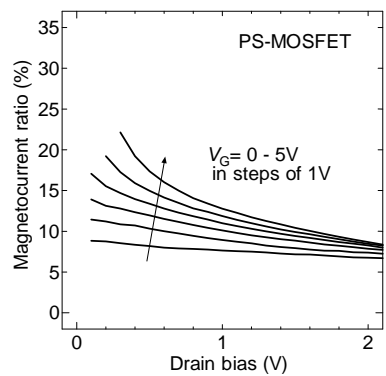
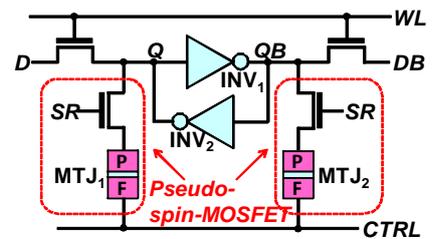

Fig. 7: Drain current as a function of magnetic field for the fabricated PS-MOSFET with $V_D$=0.5V and $V_G$=5.0V.

Fig. 8: Magnetocurrent ratio as a function of $V_D$, where $V_G$ varies from 0 to 5 V in steps of 1V.

Fig. 9: Circuit configuration of nonvolatile SRAM using PS-MOSFETs that is applicable to ideal power-gating architecture.